\documentclass[aps, prm, twocolumn, amssymb, amsmath, showpacs, superscriptaddress]{revtex4-1}

\usepackage{bm}
\usepackage{times}
\usepackage{graphicx}
\usepackage{color}
\usepackage{dcolumn}
\usepackage[colorlinks=true, letterpaper=true, pdfstartview=FitV, linkcolor=blue, citecolor=blue, urlcolor=blue]{hyperref}
\usepackage{appendix}
\usepackage[normalem]{ulem}

\begin{document}
	
	\title{Giant Anomalous Nernst Effect in Noncollinear Antiferromagnetic Mn-based Antiperovskite Nitrides}
	
	\author{Xiaodong Zhou}
	\affiliation {Key Laboratory of Advanced Optoelectronic Quantum Architecture and Measurement, Ministry of Education, School of Physics, Beijing Institute of Technology, Beijing 100081, China}
	
	\author{Jan-Philipp Hanke}
	\affiliation{Peter Gr\"unberg Institut and Institute for Advanced Simulation, Forschungszentrum J\"ulich and JARA, 52425 J\"ulich, Germany}
	
	\author{Wanxiang Feng}
	\email{wxfeng@bit.edu.cn}
	\affiliation {Key Laboratory of Advanced Optoelectronic Quantum Architecture and Measurement, Ministry of Education, School of Physics, Beijing Institute of Technology, Beijing 100081, China}
	\affiliation{Peter Gr\"unberg Institut and Institute for Advanced Simulation, Forschungszentrum J\"ulich and JARA, 52425 J\"ulich, Germany}
	
	\author{Stefan Bl\"ugel}
	\affiliation{Peter Gr\"unberg Institut and Institute for Advanced Simulation, Forschungszentrum J\"ulich and JARA, 52425 J\"ulich, Germany}
	
	\author{Yuriy Mokrousov}
	\affiliation{Peter Gr\"unberg Institut and Institute for Advanced Simulation, Forschungszentrum J\"ulich and JARA, 52425 J\"ulich, Germany}
	\affiliation{Institute of Physics, Johannes Gutenberg University Mainz, 55099 Mainz, Germany}
	
	\author{Yugui Yao}
	\affiliation {Key Laboratory of Advanced Optoelectronic Quantum Architecture and Measurement, Ministry of Education, School of Physics, Beijing Institute of Technology, Beijing 100081, China}

	\date{\today}

	\begin{abstract}
		The anomalous Nernst effect (ANE)---the generation of a transverse electric voltage by a longitudinal heat current in conducting ferromagnets or antiferromagnets---is an appealing approach for thermoelectric power generation in spin caloritronics.  The ANE in antiferromagnets is particularly convenient for the fabrication of highly efficient and densely integrated thermopiles as lateral configurations of thermoelectric modules increase the coverage of heat source without suffering from the stray fields that are intrinsic to ferromagnets.  In this work, using first-principles calculations together with a group theory analysis, we systematically investigate the spin order-dependent ANE in noncollinear antiferromagnetic Mn-based antiperovskite nitrides Mn$_{3}X$N ($X$ = Ga, Zn, Ag, and Ni).  The ANE in Mn$_{3}X$N is forbidden by symmetry in the R1 phase but amounts to its maximum value in the R3 phase.  Among all Mn$_{3}X$N compounds, Mn$_{3}$NiN presents the most significant anomalous Nernst conductivity of 1.80 AK$^{-1}$m$^{-1}$ at 200 K, which can be further enhanced if strain, electric, or magnetic fields are applied.  The ANE in Mn$_{3}$NiN, being one order of magnitude larger than that in the famous Mn$_{3}$Sn, is the largest one discovered in antiferromagnets so far. The giant ANE in Mn$_{3}$NiN originates from the sharp slope of the anomalous Hall conductivity at the Fermi energy, which can be understood well from the Mott relation. Our findings provide a novel host material for realizing \textit{antiferromagnetic spin caloritronics} which promises exciting applications in energy conversion and information processing.
	\end{abstract}

	\maketitle
	
	\section{Introduction}\label{sec1}
	Spintronics, 
	where the electron's spin degree of freedom is used as information carrier rather than its charge, has attracted enormous interest because of its promising applications in the next generation of electronic technologies.  In this context, spin-related transport phenomena arising in various magnets have been intensively investigated in the last two decades.  In ferromagnetic metals, a transverse voltage drop can be induced by a longitudinal charge current.  This phenomenon is the so-called anomalous Hall effect (AHE)~\cite{Nagaosa2010}, being one of the most competitive pathways for realizing spintronics.  Nevertheless, the energy consumption is inevitable in the AHE since the driving force has to be an external electric field.  In this light, direct coupling between spin and heat in the field of \textit{spin caloritronics}~\cite{Bauer2012,Boona2014} is more energy-efficient as spin currents can be generated by harvesting waste heat.  Therefore, spin caloritronics usually known as ``green" spintronics offers exciting prospects for energy conversion and information processing.
	
	
	The anomalous Nernst effect (ANE)~\cite{Nernst1887,D-Xiao2006,Lee2004,Weischenberg2013,L-Dong2018} $-$  the thermoelectric counterpart of the AHE $-$ is a celebrated effect from the realm of spincaloritronics.  It leads to the generation of a transverse spin-polarized charge current $j_y$ along the $y$ direction when a temperature gradient $\nabla_x T$ is applied along the $x$ direction, and therefore the ANE can be expressed as 
	\begin{equation}
		j_{y}=-\alpha_{yx}\nabla_x T,
	\end{equation}
	where $\alpha_{yx}$ is known as the anomalous Nernst conductivity (ANC).  Although ferromagnets are commonly considered as the prime sources of anomalous Nernst currents, in fact, their efficiency comes to doubt as the density of thermoelectric modules based on ferromagnets is severely limited by the effect of intrinsic stray fields from neighboring units.  Strikingly, the ANE has been recently witnessed in noncollinear antiferromagnets, such as Mn$_{3}$Sn~\cite{Ikhlas2017,XK-Li2017}, even though the net magnetization in many of such compounds vanishes.  The physics behind is the ultimate source of the ANE in  the Berry curvature of electronic states, which is promoted by breaking of proper symmetries, rather than the net magnetization itself. Since  antiferromagnets exhibit much faster dynamics than ferromagnets,  \textit{antiferromagnetic spin caloritronics}, in analogy to antiferromagnetic spintronics~\cite{Baltz2018,Zelezny2018,Jungwirth2018,Smejkal2018}, is becoming an attractive research field.  The ANC in Mn$_{3}$Sn is considerably larger than that in 3d transition-metal ferromagnets (e.g., Fe and Co)~\cite{Ikhlas2017,XK-Li2017}, while it is still one order of magnitude smaller than that in the full-Heusler ferromagnet Co$_{2}$MnGa, which exhibits the largest ANC reported to date~\cite{Sakai2018,Guin2019}.  Since the ANC is sensitive to the details of the electronic structure for a given magnetic material, finding antiferromagnets which host large ANE is a crucial step to realize antiferromagnetic spin caloritronics.
	
	In addition to Mn$_{3}$Sn, the antiperovskite Mn$_{3}X$N ($X$ = Ga, Zn, Ag, Ni, etc.) presents another important class of noncollinear antiferromagnets, which was known since the 1970s~\cite{Bertaut1968,Fruchart1978}.  Many unique physical properties have been found in Mn$_{3}X$N, including the magnetovolume effects~\cite{Gomonaj1989,Gomonaj1992,WS-Kim2003, Lukashev2008,Lukashev2010,Takenaka2014,SH-Deng2015,Zemen2017a}, magnetocaloric effects~\cite{Y-Sun2012,Matsunami2014,KW-Shi2016,Zemen2017}, magneto-optical effect~\cite{XD-Zhou2019} and AHE~\cite{XD-Zhou2019,Gurung2019,K-Zhao2019,Huyen2019}.  However, the ANE, being a practical scheme for spin caloritronics~\cite{Bauer2012,Boona2014}, has not been reported till now in this class of materials.  This motivated us to explore the ANE in Mn$_{3}X$N in order to find a superior antiferromagnetic host material which couples spin transport with heat most efficiently.
	
\begin{figure*}
		\centering
		\includegraphics[width=1.8\columnwidth]{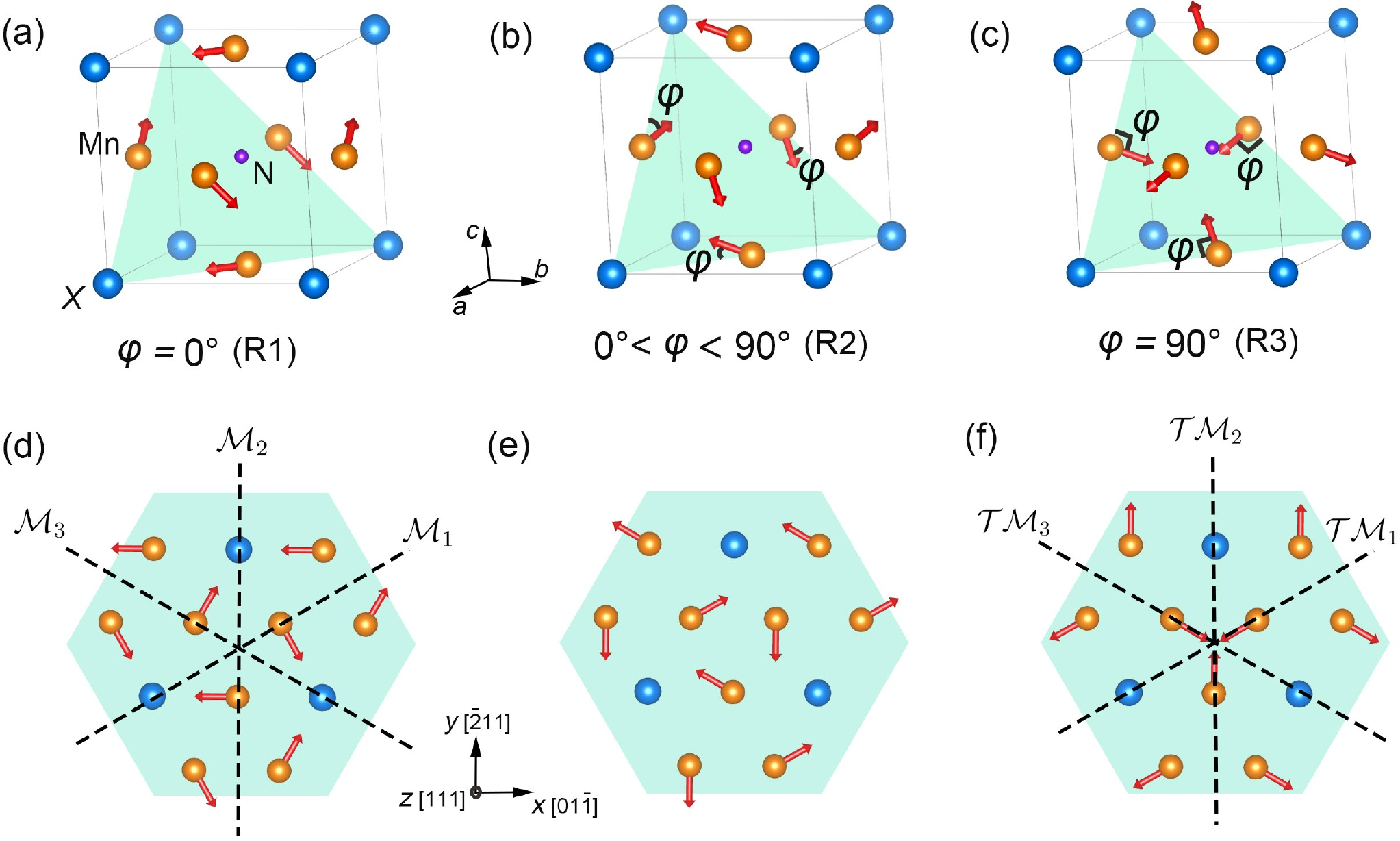}
		\caption{(Color online) Crystal and magnetic structures of Mn$_{3}X$N.  (a)--(c) Different spin orders in Mn$_3X$N as classified by the R1, R2, and R3 phases.  The yellow, blue, and purple spheres represent Mn, $X$, and N atoms, respectively.  The red arrows label the directions of the spin magnetic moments of Mn atoms, which lie on the (111) plane highlighted in green.  The azimuthal angle $\varphi$, defined as the spin order parameter, measures the rotation of spins away from the face diagonals of the cube.  (d)--(f) Top view of the noncollinear spin order in the (111) plane.  $\mathcal{M}_{1}$, $\mathcal{M}_{2}$, and $\mathcal{M}_{3}$ in (d) are three mirror symmetries; $\mathcal{TM}_{1}$, $\mathcal{TM}_{2}$, and $\mathcal{TM}_{3}$ in (f) are the symmetries combining mirror and time-reversal ($\mathcal{T}$) operations.  The $\mathcal{C}_{3}$ rotation with respect to the $z$ axis is always present in (d)--(f).}
		\label{fig1}
	\end{figure*}

	In this work, using state-of-the-art first-principles calculations, we systematically study the ANE in noncollinear antiferromagnetic antiperovskite Mn$_{3}X$N ($X$ = Ga, Zn, Ag, and Ni).  We first show that the ANE depends strongly on the spin order, which characterizes the 120$^\circ$ noncollinear spin structure (Fig.~\ref{fig1}). Using group theory analysis and Berry curvature calculations, we demonstrate that the ANE in Mn$_{3}X$N vanishes in the R1 phase while it assumes its maximal value in the R3 phase.  The system Mn$_{3}$NiN has an ANC that is as large as $\sim$2 AK$^{-1}$m$^{-1}$, which is nearly one order of magnitude larger than that in the noncollinear antiferromagnet Mn$_{3}$Sn ($\sim$0.2 AK$^{-1}$m$^{-1}$)~\cite{Ikhlas2017,XK-Li2017} and is close to the reported largest ANC in the ferromagnet Co$_{2}$MnGa ($\sim$4.0 AK$^{-1}$m$^{-1}$)~\cite{Sakai2018,Guin2019}.  The pronounced ANC in Mn$_{3}$NiN originates from the steep slope of the anomalous Hall conductivity (AHC) at the Fermi energy, and can be understood from the Mott relation.  Moreover, the ANE in Mn$_{3}$NiN can be tuned by strain, electric, and magnetic fields.  This pronounced and tunable ANE suggests Mn$_3$NiN as an ideal material platform for realizing highly efficient spin thermoelectric devices based on noncollinear antiferromagnets rather than traditional ferromagents as schematically shown in Fig.~\ref{fig4}.
	
\section{Result and Discussion}\label{sec2}	
	
	Mn$_{3}X$N forms an antiperovskite crystal structure that hosts a noncollinear antiferromagnetic order, as displayed in Fig.~\ref{fig1}.  The nonmagnetic $X$ and N atoms occupy the corners and the center of the cube, respectively, whereas the three magnetic Mn atoms are located on the face centers.  The spin magnetic moments of Mn atoms lie on the (111) plane and form a 120$^{\circ}$ noncollinear spin structure.  Interestingly, all three spins can simultaneously rotate within the (111) plane, depending on the temperature $T$~\cite{Bertaut1968,Fruchart1978}.  In order to quantitatively describe the noncollinear spin structure, we introduce the azimuthal angle $\varphi$ as spin order parameter, measuring the rotation of the spins away from the face diagonals of the cube.  Taking Mn$_{3}$NiN as an example~\cite{Fruchart1978}, the three spins align along the face diagonals of the cube for $T<$ 163 K, which is called R1 phase ($\varphi=0^{\circ}$) (Figs.~\ref{fig1}a and~\ref{fig1}d).  When 163 K $<T<$ 266 K, the R3 phase ($\varphi=90^{\circ}$) can appear (Figs.~\ref{fig1}c and~\ref{fig1}f), for which the three spins point to the center of the triangle formed by neighboring magnetic atoms.  The R1 and R3 phases are also called $\Gamma^{5g}$ and $\Gamma^{4g}$ spin configurations, respectively~\cite{Bertaut1968,Fruchart1978}.  An intermediate R2 phase ($0^{\circ} < \varphi < 90^{\circ}$) (Figs.~\ref{fig1}b and~\ref{fig1}e) is expected to emerge during the phase transition between the R1 and R3 phases~\cite{Gomonaj1989,Gomonaj1992}.  Such temperature-dependent noncollinear spin structure may also be realized in other Mn$_{3}X$N ($X$ = Ag, Ga, Zn) compounds~\cite{Bertaut1968,Fruchart1978,Gomonaj1989,Gomonaj1992}.
	
	The R3 phase of Mn$_{3}X$N could have the weak ferromagnetism along the crystallographic [111] direction,  and the calculated total spin magnetic moment is only 0.006 $\mu_{B}$ (0.002 $\mu_{B}$ per Mn site).  The weak spin ferromagnetism is not responsible for the emergence of the ANE, while the noncollinear spin order is the ultimate source as we demonstrate later by using group theory.  In practice, the directions of spin magnetic moments are constrained within the (111) plane such that we actually consider a fully compensated antiferromagnet which has vanishing total spin magnetization.  Our calculations also show that the total orbital magnetic moment is not vanishing; however, one cannot merely say that it induces the ANE in Mn$_{3}X$N.  The orbital magnetization is very closely related to the AHE and ANE via the Berry curvature of the electronic states~\cite{D-Xiao2006,L-Dong2018}.  It is the symmetry properties of the Berry curvature which are perceived as the main origin of the emergence of the orbital magnetization, AHE, or ANE in both ferromagnetic and antiferromagnetic materials.  The orbital ferromagnetism arises hand in hand with the ANE since both are allowed by symmetry, and it would be improper to argue that orbital magnetism is the origin of the effects discussed in Mn$_{3}X$N.  It would be different from the case of Ref.~\cite{Solovyev1997}, in which it is the combination of the antiferromagnetic order with lowered crystal symmetry and spin-orbit interaction which breaks the necessary symmetry for the emergence of the orbital magnetization and magneto-optical effects.
	
	\begin{table}[b!]
		\centering
		\caption{Magnetic point groups and symmetry-allowed elements of the anomalous Nernst conductivity (ANC) tensor for Mn$_{3}X$N as a function of the azimuthal angle $\varphi$ that defines the noncollinear spin order.  The magnetic point groups are calculated by the \textsc{isotropy} code~\cite{Stokes}.}
		\label{tab1}
		\setlength{\tabcolsep}{3pt} 
		\renewcommand{\arraystretch}{1.5} 
		\begin{tabular}{cccccccc}
			\hline
			Azimuthal angle $\varphi$ & $0^{\circ}$&$30^{\circ}$&$60^{\circ}$&$90^{\circ}$&$120^{\circ}$&$150^{\circ}$&$180^{\circ}$ \\
			\hline
			Magnetic point group & $\bar{3}1m$ & $\bar{3}$ & $\bar{3}$ & $\bar{3}1m^{\prime}$ & $\bar{3}$ & $\bar{3}$ & $\bar{3}1m$ \\
			Nonzero ANC element & -- & $\alpha_{xy}$ & $\alpha_{xy}$ & $\alpha_{xy}$ &  $\alpha_{xy}$ & $\alpha_{xy}$ & -- \\
			\hline
		\end{tabular}
	\end{table}
	
	The ANE in Mn$_{3}X$N can be anticipated to significantly depend on the spin order.  To demonstrate this, we first employ group theory to analyze the influence of the different spin orders on the Berry curvature, which is the key quantity in calculating the ANC [see Eqs.~\eqref{eq:IANC} and~\eqref{eq:BerryCur}].  Since the ANC is a pseudovector, it can be written in a vector notation, $\boldsymbol{\alpha}=[\bar{\alpha}^{x},\bar{\alpha}^{y},\bar{\alpha}^{z}]\equiv[\alpha_{yz},\alpha_{zx},\alpha_{xy}]$, where the vector components correspond one-by-one to the off-diagonal elements of the Nernst conductivity tensor, i.e., $\bar{\alpha}^{x/y/z}\equiv\alpha_{yz/zx/xy}$.  Similarly, the Berry curvature can be written as $\boldsymbol{\Omega}_{n}=[\bar{\Omega}_{n}^{x},\bar{\Omega}_{n}^{y},\bar{\Omega}_{n}^{z}]\equiv[\Omega^{n}_{yz},\Omega^{n}_{zx},\Omega^{n}_{xy}]$, where $n$ is the band index.  Both properties are translationally invariant such that it is sufficient to restrict our analysis to magnetic point groups.  Table~\ref{tab1} lists the evolution of the magnetic point group with the spin order parameter $\varphi$.  One can see that the magnetic point group exhibits a period of $\pi$ and there are three non-repetitive elements, $\bar{3}1m$ [$\varphi=n\pi$], $\bar{3}1m^{\prime}$ [$\varphi=(n +\frac{1}{2})\pi$], and $\bar{3}$ [$\varphi \neq n\pi$ and $\varphi \neq (n +\frac{1}{2})\pi$] with $n \in \mathbb{N}$, which we shall discuss one by one.  First, $\bar{3}1m$ contains three mirror planes: $\mathcal{M}_{1}$, $\mathcal{M}_{2}$, and $\mathcal{M}_{3}$ (Fig.~\ref{fig1}d).  $\mathcal{M}_{2}$ is parallel to the $yz$ plane, which changes the sign of $\bar{\Omega}_{n}^{y}$ and $\bar{\Omega}_{n}^{z}$ but preserves $\bar{\Omega}_{n}^{x}$.  This implies that $\bar{\Omega}_{n}^{y}$ and $\bar{\Omega}_{n}^{z}$ are odd functions of $k_{x}$, while $\bar{\Omega}_{n}^{x}$ is an even function.  By integrating the Berry curvature over the entire Brillouin zone, we arrive at $\boldsymbol{\alpha}=[\bar{\alpha}^{x},0,0]$.  In addition, $\bar{3}1m$ contains a three-fold rotation symmetry $\mathcal C_{3}$ around the [111] direction that relates $\mathcal{M}_{2}$ to the other two mirror planes $\mathcal{M}_{1}$ and $\mathcal{M}_{3}$.  Since any component of the ANC normal to the $\mathcal{C}_{3}$ axis, for example $\bar{\alpha}^{x}$, is forced to be zero, it finally results in $\boldsymbol{\alpha}=[0,0,0]$ under the group $\bar{3}1m$.  Therefore, the ANE is forbidden by symmetry in the R1 phase ($\varphi=n\pi$).  Second, in contrast to $\bar{3}1m$, all mirror planes are absent in the group $\bar{3}$ and only the $\mathcal{C}_3$ axis is preserved (Fig.~\ref{fig1}e).  This leads to vanishing $\bar{\Omega}_{n}^{x}$ and $\bar{\Omega}_{n}^{y}$, and there exists $\boldsymbol{\alpha}=[0,0,\bar{\alpha}^{z}]=[0,0,\alpha_{xy}]$ in the R2 phase [$\varphi \neq n\pi$ and $\varphi \neq (n +\frac{1}{2})\pi$].  Third, $\bar{3}1m^{\prime}$ contains operations combining time and space symmetries: $\mathcal{TM}_{1}$, $\mathcal{TM}_{2}$, and $\mathcal{TM}_{3}$ (Fig.~\ref{fig1}f).  As mentioned above, $\bar{\Omega}_{n}^{y}$ and $\bar{\Omega}_{n}^{z}$ are odd but $\bar{\Omega}_{n}^{x}$ is even with respect to $\mathcal{M}_{2}$.  By considering further that all components $\bar{\Omega}_{n}^{i}$ are odd under the time-reversal operation $\mathcal{T}$, we find that $\bar{\Omega}_{n}^{y}$ and $\bar{\Omega}_{n}^{z}$ are even under $\mathcal{TM}_2$ whereas $\bar{\Omega}_{n}^{x}$ is odd, giving rise to $\boldsymbol{\alpha}=[0,\bar{\alpha}^{y},\bar{\alpha}^{z}]$.  Since $\bar{\alpha}^{y}$ is forced to be zero due to the $\mathcal{C}_{3}$ operation, only $\bar{\alpha}^{z}$ is nonzero in the R3 phase [$\varphi=(n +\frac{1}{2})\pi$].  The symmetry-allowed ANC elements and the corresponding magnetic point groups are summarized in Table~\ref{tab1}.
	
	\begin{figure*}
		\centering
		\includegraphics[width=2.0\columnwidth]{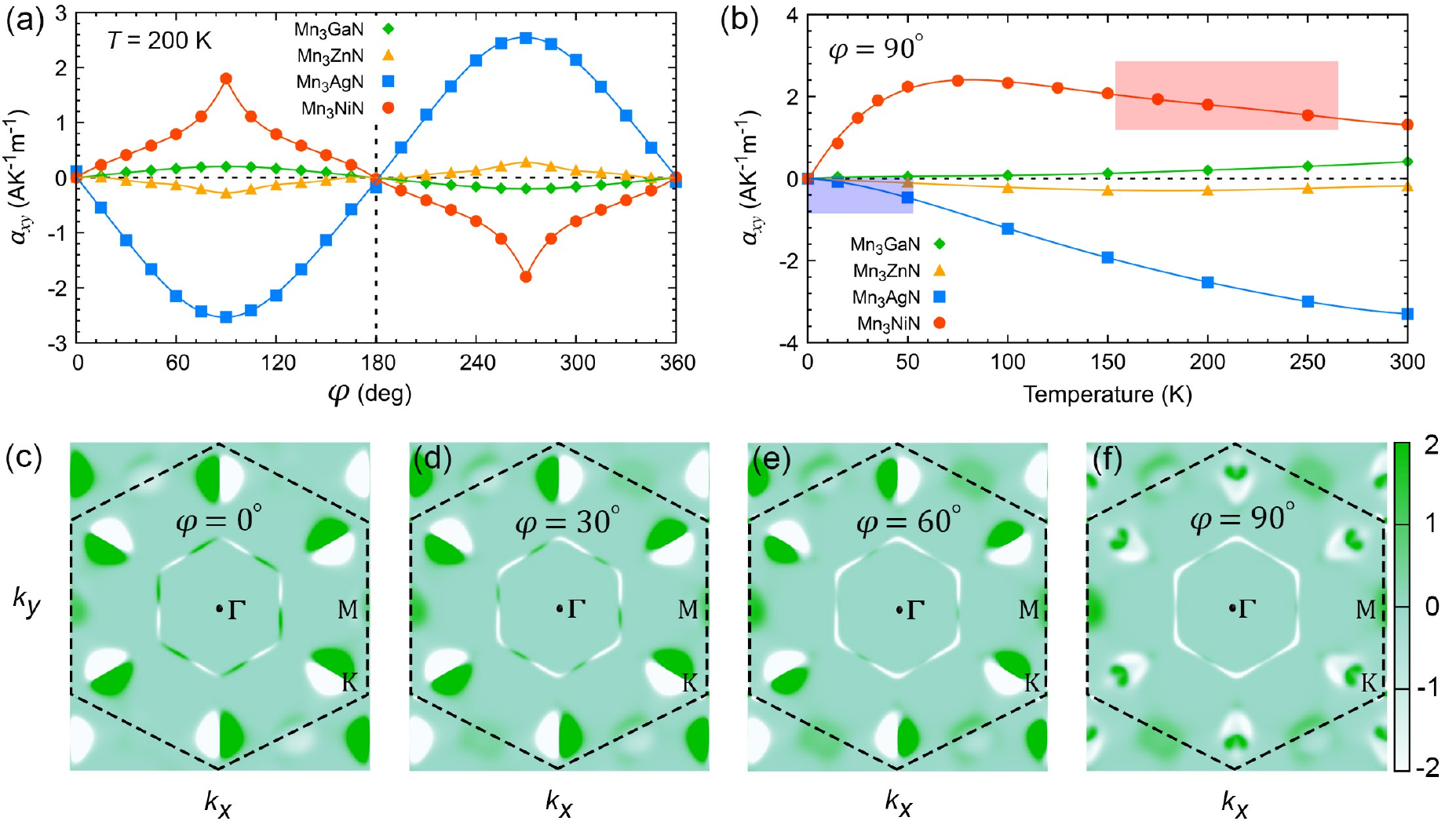}
		\caption{(Color online) Anomalous Nernst conductivity (ANC) and Berry curvature in Mn$_{3}X$N.  (a) ANC as a function of the azimuthal angle $\varphi$ at the temperature $T$ = 200 K.  (b) ANC as a function of temperature when $\varphi=90^\circ$ (R3 phase), where the shaded regions indicate the temperature regimes in which the R3 phase exists in Mn$_{3}$NiN and Mn$_{3}$AgN.  The solid lines in (a) and (b) are the polynomial fittings.  (c)--(f) The Berry curvature (in arbitrary units) of Mn$_{3}$NiN on the $k_{z} = 0$ plane when $T$=200 K for the azimuthal angles of $\varphi$ = 0$^\circ$, 30$^\circ$, 60$^\circ$, and 90$^\circ$, respectively.  The dashed black lines mark the first Brillouin zone.}
		\label{fig2}
	\end{figure*}
	
	Although group theory is particularly powerful to identify the shape of the ANC tensor, it does not help us to evaluate the magnitude of the symmetry-allowed elements of the ANC, which are sensitive to details of the electronic structure.  In the following, first-principles calculations are used as a quantitative method to predict the ANE in Mn$_{3}X$N.  Fig.~\ref{fig2}a presents the intrinsic ANC as a function of the spin order parameter $\varphi$ at a temperature of 200 K.  The ANC vanishes when $\varphi=n\pi$ but turns out to be finite if $\varphi\neq n\pi$, which is in full accordance with the above symmetry arguments.  Nevertheless, the ANC in Mn$_{3}X$N displays a curve that has a period of $2\pi$ in $\varphi$ and gives rise to the maxima at $\varphi=(n +\frac{1}{2})\pi$.  In order to understand this observation, we evaluate the total Berry curvature as the weighted sum $\Omega_{xy}(\bm{k})=\sum_{n}W_{n}(\bm{k})\Omega^{n}_{xy}(\bm{k})$ over all bands with weights $W_n$ given by Eqs.~\eqref{eq:WnT}.  Figs.~\ref{fig2}c--\ref{fig2}f show the resulting momentum-space distribution in the $k_z=0$ plane.  One can see that in the R1 phase ($\varphi=0^\circ$) the symmetrically distributed hot spots, which have same magnitude but opposite sign, cancel out each other, leading overall to a vanishing ANC.  In the R2 phase (e.g., $\varphi=30^\circ$ and $60^\circ$), however, the distribution of these hot spots becomes more asymmetric with increasing $\varphi$.  Eventually, in the R3 phase with $\varphi=90^\circ$, the difference between the positive and negative microscopic contributions reaches a maximum, manifesting in the largest ANC.  Additionally, as the ANC inherits its symmetry properties from the Berry curvature, the ANC follows the relation $\alpha_{xy}(\varphi) = -\alpha_{xy}(\varphi+\pi)$, which discloses that the spin order at $\varphi+\pi$ is the time-reversed counterpart of the one at $\varphi$, and the ANC is odd under time-reversal symmetry.  Since the R3 phases with $\varphi=90^\circ$ and $\varphi=270^\circ$ have the same absolute value but opposite sign of $\alpha_{xy}$, they can be naturally chosen as two neighboring thermoelectric modules, in which the directions of electric fields reverse, without suffering from any of the obstructive stray fields known from ferromagnetic thermopiles (cf. Figs.~\ref{fig4}b and~\ref{fig4}c).
	
	\begin{figure*}
		\centering
		\includegraphics[width=2.0\columnwidth]{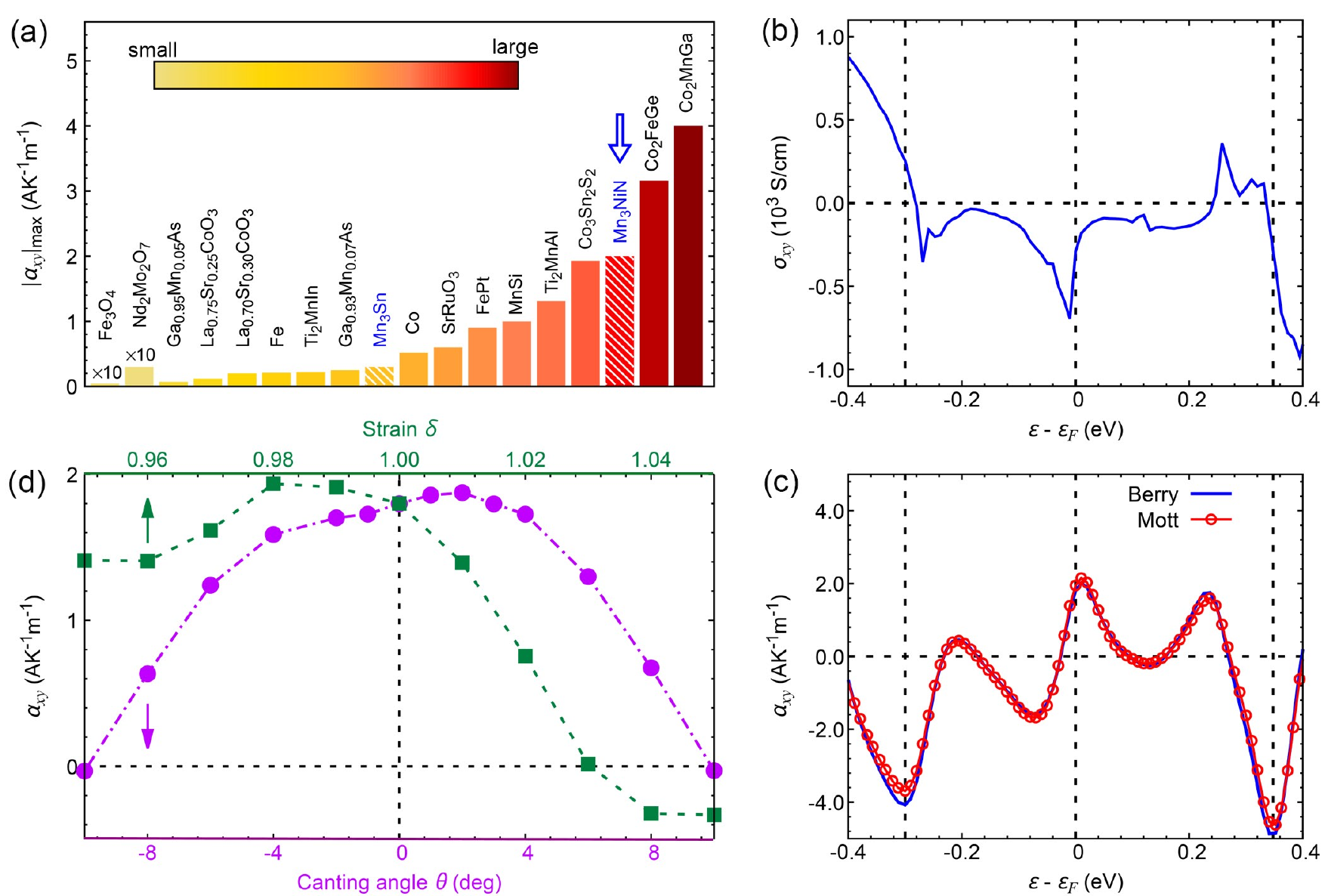}
		\caption{(Color online) Origin of the large ANE in Mn$_{3}$NiN and its tunability by various fields.  (a) The ANC recorded for various ferromagnets and two antiferromagnets (Mn$_{3}$Sn and Mn$_{3}$NiN).  Part of the data is taken from the previous works.~\cite{Sakai2018,Guin2019a,Noky2018,Noky2018a}  (b)--(c) The AHC and ANC of Mn$_{3}$NiN as a function of energy.  The ANC $\alpha_{xy}$ is calculated at the temperature $T=200$ K using the formula of Berry curvature [Eqs.~\eqref{eq:IANC}] and also the generalized Mott relation [Eqs.~\eqref{eq:Mott1}].  (d) The ANC $\alpha_{xy}$ of Mn$_{3}$NiN as a function of canting angle $\theta$ and strain $\delta$ at $T=200$ K.  The positive and negative values of $\theta$ indicate the canting of all three spins along the $[111]$ and $[\bar{1}\bar{1}\bar{1}]$ directions, respectively.}
		\label{fig3}
	\end{figure*}
	
	Being a thermal transport phenomenon, the ANE should rely substantially on the actual temperature.  Fig.~\ref{fig2}b shows how temperature influences the ANE in the R3 phase of Mn$_{3}X$N. Mn$_{3}$GaN and Mn$_{3}$ZnN are not appealing since their ANC is much smaller than that of Mn$_{3}$NiN and Mn$_3$AgN below 300 K.  The ANC in Mn$_{3}$AgN increases monotonically with increasing temperature and it exceeds the value in Mn$_{3}$NiN for $T>$ 150 K, however, the N{\'e}el temperature of Mn$_{3}$AgN of below 55 K is rather low~\cite{Fruchart1978}.  The R3 phase of Mn$_3$NiN is realized in a broad range of temperatures from 163 K to 266 K~\cite{Fruchart1978}, in which the ANC varies from 1.98 AK$^{-1}$m$^{-1}$ to 1.47 AK$^{-1}$m$^{-1}$.  The ANC in Mn$_{3}$NiN is nearly one order of magnitude larger than that in noncollinear antiferromagnetic Mn$_{3}$Sn ($\sim$0.2 AK$^{-1}$m$^{-1}$).  The pronounced ANC that we predict for Mn$_{3}$NiN is substantially larger than for most of the typical ferromagnets (0.01$\sim$1 AK$^{-1}$m$^{-1}$), and the calculated value is only slightly smaller than for the two ferromagnetic Weyl semimetals Co$_{2}$FeGe~\cite{Noky2018} and Co$_{2}$MnGa~\cite{Sakai2018,Guin2019}, as summarized in Fig.~\ref{fig3}a.  Here, we stress that Co$_{2}$FeGe and Co$_{2}$MnGa as intrinsic ferromagnets do not play any role for antiferromagnetic spin caloritronics as they are not free of parasitic stray fields.
	
	Next, we demonstrate the underlying physical mechanism of the large ANC in Mn$_{3}$NiN by relating the ANC $\alpha_{xy}$ to the anomalous Hall conductivity (AHC) $\sigma_{xy}$ via the generalized Mott formula~\cite{D-Xiao2006}:
	\begin{equation}\label{eq:Mott1}
		\alpha_{xy}=-\frac{1}{e}\int d\varepsilon\frac{\partial f}{\partial \mu}\sigma_{xy}\frac{\varepsilon-\mu}{T},
	\end{equation}
	where $e$ is the elementary positive charge, $\varepsilon$ is the energy, $\mu$ is the chemical potential of the electrons, and $f(\varepsilon)=1/[\text{exp}((\varepsilon-\mu)/k_{B}T)+1]$ is the Fermi-Dirac distribution function.  In the zero temperature limit, the integral in Eqs.~\eqref{eq:Mott1} can be carried out by the Sommerfeld expansion to the lowest order term~\cite{Ashcroft1976}.  Then, the standard Mott formula, which relates the ANC to the energy derivative of the AHC, is obtained:
	\begin{equation}\label{eq:Mott2}
		\alpha_{xy} = -\dfrac{\pi^{2}k_{B}^{2}T}{3e}\left.\frac{d\sigma_{xy}}{d\varepsilon}\right|_{\varepsilon=\mu}.
	\end{equation}
	As can be seen from Eqs.~\eqref{eq:Mott2}, we can expect a large ANC in a given system if the corresponding AHC changes rapidly with energy at the Fermi level for $\mu=\varepsilon_F$. Figs.~\ref{fig3}b and~\ref{fig3}c present the variation of $\sigma_{xy}$ and $\alpha_{xy}$ in Mn$_3$NiN as a function of the energy $\varepsilon$, respectively.  The AHC amounts to a moderate value of $\sigma_{xy}(\varepsilon_{F})=291$ S/cm, but the slope of the curve at $\varepsilon_{F}$ is very large.  It thus results in a prominent ANC of $\alpha_{xy}(\varepsilon_{F})=1.80$ AK$^{-1}$m$^{-1}$.  Furthermore, $\alpha_{xy}$ can increase up to 2.0 AK$^{-1}$m$^{-1}$ by slightly moving $\varepsilon_{F}$ upward by 0.01 eV, which could be easily realized by electron doping, e.g., in the alloy Mn$_{3}$Ni$_{1-x}$Cu$_{x}$N~\cite{K-Zhao2019}.  If a relatively heavy doping concentration is achieved, the ANC can reach up to -4.08 AK$^{-1}$m$^{-1}$ at -0.30 eV and even up to -4.87 AK$^{-1}$m$^{-1}$ at 0.34 eV, the latter exceeding the ANC in ferromagnetic Co$_{2}$MnGa ($\sim$4.0 AK$^{-1}$m$^{-1}$)~\cite{Sakai2018,Guin2019}.  Overall, the origin of the prominent ANC in Mn$_{3}$NiN is rooted in the large energy derivative of the AHC at the Fermi level in accordance with the Mott relation.
	
	\begin{figure*}
		\centering
		\includegraphics[width=2.0\columnwidth]{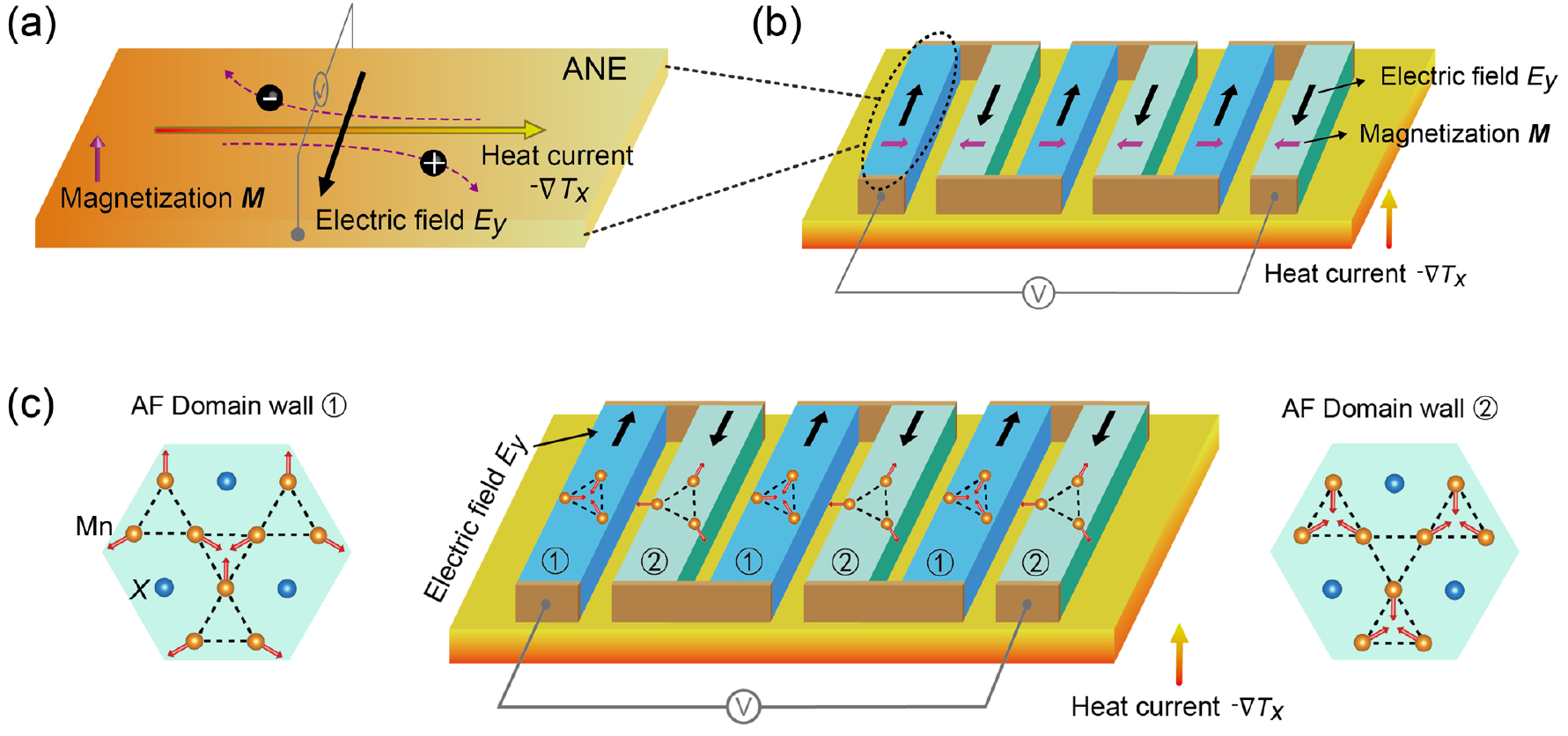}
		\caption{(Color online) Schematics of the anomalous Nernst effect and thermopile structures.  (a) Anomalous Nernst effects (ANE).  (b) Thermopile made out of an array of thermoelectric modules that use the ANE in collinear ferromagnets.  (c) Center panel: Unconventional thermopile based on the ANE in chiral noncollinear antiferromagnets;  Left and right panels: The two noncollinear antiferromagnetic domain walls, as seen from the (111) plane of cubic Mn$_{3}X$N, are mutual partners connected by time-reversal symmetry.}
		\label{fig4}
	\end{figure*}

	In the light of practical applications of antiferromagnetic spin caloritronics, it is particularly important to understand how to control and design the ANE in Mn$_3X$N by various external means, including strain, electric, and magnetic fields.  First, a strain field could become active due to the lattice mismatch between the thin film of the sample and a substrate, for example, in Mn$_{3}$NiN~\cite{Boldrin2019a,Boldrin2019} and another  noncollinear antiferromagnet Mn$_{3}$Pt~\cite{ZQ-Liu2018}.  Since the AHE can be effectively tuned by strain fields~\cite{Boldrin2019,ZQ-Liu2018}, we anticipate that the ANE is susceptible in a similar way.  Here, we consider strain along the [111] direction, quantified by $\delta = d/d_0$, where $d$ and $d_{0}$ are the distances between two neighboring (111) planes in the strained and unstrained cases, respectively.  Considering the Poisson effect, the lattice within the (111) plane should shrink (expand) when $\delta>1$ ($\delta<1$) and the constant volume approximation is used.  The ANC as a function of $\delta$ is displayed in Fig.~\ref{fig3}d, in which one can see that tensile strain suppresses $\alpha_{xy}$ and even changes its sign, while compressive strain leads to a larger $\alpha_{xy}$ for $0.98<\delta<1.0$.  If the thin film grows epitaxially along the (111) direction, substrates with larger lattice constants would be beneficial to generate a larger ANC.  Second, the magnitude of ANC could be altered by an electric field since the Fermi energy will be shifted if a gate voltage is applied.  Fig.~\ref{fig3}c clearly shows that the ANC depends on the position of the Fermi energy.  The role of gate voltage can be replaced by alloying (e.g., Mn$_{3}$Ni$_{1-x}$Cu$_{x}$N~\cite{K-Zhao2019}) which also introduces electron or hole doping.  Third, as proposed in Mn$_{3}$Sn~\cite{Rout2019}, an external magnetic field normal to the (111) plane can induce an out-of-plane spin canting (i.e., pointing to the [111] direction) to form a noncoplanar spin structure, which is responsible for emergence of the topological Hall effect. Interestingly, hydrostatic pressure plays a similar role like the magnetic field, and even induces a larger spin canting angle in Mn$_{3}$Ge~\cite{Sukhanov2018}.  Fig.~\ref{fig3}d shows that $\alpha_{xy}$ increases with the increasing of the canting angle $\theta$ from $0^{\circ}$ to $2^{\circ}$, while $\alpha_{xy}$ reduces rapidly to zero when $|\theta|>4^\circ$.  Following the same rationale as for the tunability of the AHE in noncollinear antiferromagnets~\cite{Boldrin2019,ZQ-Liu2018,Rout2019,Sukhanov2018}, we thereby demonstrated that the prominent ANE in Mn$_3$NiN can be controlled by various external fields, offering great prospects for energy-efficient applications based on antiferromagnetic spin caloritronics.

	Finally, we design a thermopile structure based on noncollinear antiferromagnets (e.g., Mn$_{3}$NiN) and compare it with the conventional ones composed of collinear ferromagnets (Fig~\ref{fig4}).  Fig.~\ref{fig4}a depicts the basic principle of the ANE, that is, a transverse charge current is generated by a longitudinal thermal current and both of these currents are perpendicular to the direction of magnetization.  Exhibiting the ANE, collinear ferromagnets are usually made into a thermopile in which the directions of magnetization in neighboring thermoelectric modules have to be opposite to form an electrical circuit (Fig.~\ref{fig4}b).  However, this obstructs the miniaturization of the devices as the density of thermoelectric modules is severely limited by the inherent stray fields in neighboring modules.  This issue can be solved if noncollinear antiferromagnets are used instead, as sketched in Fig.~\ref{fig4}c.  The antiferromagnetic domain walls with inverted spin patterns, in which the charge currents flow into opposite directions (cf. Fig.~\ref{fig2}a), can be arranged much closer to maximize the coverage of heat source without suffering from parasitic stray field.  Consequently, the antiferromagnetic thermopile structure is superior to the conventional ferromagnetic one, uncovering the bright prospects of \textit{antiferromagnetic spin caloritronics} for materials science and devices physics.

\section{Summary}\label{sec3}
	
	In conclusion, employing first-principles calculations together with a group theory analysis, we investigated the spin order-dependent ANE in noncollinear antiferromagnets Mn$_{3}X$N with $X$ = Ga, Zn, Ag, and Ni.  By using group theory, we uncovered that the ANE can emerge in Mn$_{3}X$N, except for the R1 phase characterized by the spin order parameter $\varphi=n\pi$.  The first-principles calculations supported the group theory analysis and further revealed that the R3 phase [$\varphi=(n+\frac{1}{2})\pi$] has the largest ANC $\alpha_{xy}$.  The asymmetrical distribution of the hot spots of Berry curvature explained well the variation of $\alpha_{xy}$ with $\varphi$.  Mn$_{3}$NiN was identified to be the most interesting material among all four Mn$_{3}X$N compounds because its noncollinear state exists over a broad range of temperatures (163 K to 266 K), for which $\alpha_{xy}$ amounts to as much as 1.98 AK$^{-1}$m$^{-1}$.  The giant ANC in Mn$_{3}$NiN originated from a pronounced energy variation of the AHC at the Fermi level, which can be well understood by the Mott relation.  Moreover, we demonstrated that both magnitude and sign of the ANE can be controlled by designing external perturbations in terms of strain, electric field, or magnetic field.  It should be stressed that the ANC in Mn$_{3}$NiN is one order of magnitude larger than that in the famous noncollinear antiferromagnet Mn$_{3}$Sn.  Thus, our results promote the chiral magnet Mn$_{3}$NiN as an ideal material platform for establishing antiferromagnetic spin caloritronics as an intriguing pathway for energy conversion and information processing.

\begin{acknowledgments}
	The authors thank helpful discussion with Guang-Yu Guo.  W.F. and Y.Y. acknowledge the support from the National Natural Science Foundation of China (Nos. 11874085, 11734003, and 11574029) and the National Key R\&D Program of China (No. 2016YFA0300600).  W.F. also acknowledges the funding through an Alexander von Humboldt Fellowship. X.Z. thanks the supports from Graduate Technological Innovation Project of Beijing Institute of Technology (Grant No. 2019CX10018).  Y.M. and S.B. acknowledge the funding under SPP 2137 ``Skyrmionics" (project MO 1731/7-1), collaborative Research Center SFB 1238, and Y.M. acknowledges the funding from project MO 1731/5-1 and SFB/TRR 173 of Deutsche Forschungsgemeinschaft (DFG). We also gratefully acknowledge the J{\"u}lich Supercomputing Centre and JARA-HPC of RWTH Aachen University for providing computational resources under project jiff40.
\end{acknowledgments}

\begin{table}[ht!]
	\centering
	\caption{The calculated anomalous Nernst conductivities, $|\alpha_{xy}|$ or $|\alpha_{yz}|$ (AK$^{-1}$m$^{-1}$), in a traditional ferromagnet (bcc Fe), ferromagnetic Weyl semimetals (Co$_{2}$FeGe and Co$_{2}$MnGa), compensated ferrimagnets (Ti$_{2}$MnAl and Ti$_{2}$MnIn), and a noncollinear antiferromagnet (Mn$_{3}$Sn).  To compare the experimental value of Co$_{2}$MnGa, the Fermi energy has been shifted to the energy $\varepsilon=\varepsilon_{F}+0.07$ eV and $\varepsilon=\varepsilon_{F}+0.08$ eV in the present and previous~\cite{Guin2019} works, respectively.}
	\label{tab2}
	\setlength{\tabcolsep}{1.5pt} 
	\renewcommand{\arraystretch}{1}
	\begin{tabular*}{\columnwidth}{@{\extracolsep{\fill}}lllllll}
		\hline
		& Fe & Co$_2$FeGe & Co$_2$MnGa & Ti$_2$MnAl & Ti$_2$MnIn   & Mn$_3$Sn \\
		& (300 K) & (300 K) & (300 K) & (300 K) & (300 K) & (200 K) \\
		\hline				
		$|\alpha_{xy}|$ & 0.49 & 3.46 & 4.01 &1.24 &0.29   & 0.25 \\
		& 0.40\footnotemark[1] & 3.16\footnotemark[2]  & 4.00\footnotemark[3]  & 1.31\footnotemark[4]
		& 0.22\footnotemark[4]  & 0.28\footnotemark[5] \\
		\hline
	\end{tabular*}
	\footnotemark[1]{Ref. [\citenum{GY-Guo2017}] (theory),}
	\footnotemark[2]{Ref. [\citenum{Noky2018}] (theory),}
	\footnotemark[3]{Refs. [\citenum{Sakai2018,Guin2019}] (theory and experiment),}
	\footnotemark[4]{Ref. [\citenum{Noky2018a}] (theory),}
	\footnotemark[5]{Ref. [\citenum{Ikhlas2017}] (experiment).}
\end{table}
	
	\appendix
	\section{The details of first-principles calculations}\label{appendix}
	
	First-principles calculations were performed using the projector augmented wave method~\cite{Bloechl1994} as implemented in the Vienna \textit{ab initio} simulation package~\cite{Kresse1996a}.  The exchange-correlation functional was treated by the generalized-gradient approximation with the Perdew-Burke-Ernzerhof parameterization~\cite{Perdew1996}.  The lattice constants of Mn$_3X$N ($X$ = Ga, Zn, Ag, and Ni) were adopted to the experimental values of 3.898, 3.890, 4.013, and 3.886 {\AA}, respectively~\cite{Takenaka2014}.  Spin-orbit coupling was included in all calculations, the energy cut-off was chosen as 500 eV, the energy criterion was 10$^{-6}$ eV, and a $k$-mesh of 16$\times$16$\times$16 points was used.  During the self-consistent field calculations, a penalty functional was added into the total-energy expression to constrain the direction of the spin magnetic moments within the (111) plane.  After obtaining the converged charge density, we constructed maximally localized Wannier functions by projecting onto $s$, $p$, and $d$ orbitals of Mn and $X$ atoms as well as onto $s$ and $p$ orbitals of N atom, using a uniform $k$-mesh of 10$\times$10$\times$10 points in conjuction with the \textsc{wannier90} package~\cite{Mostofi2008}.  Then, transverse electronic and thermoelectric transport properties were calculated using the accurate \textit{ab initio} tight-binding Hamiltonian on the basis of Wannier functions.
	
	Taking advantage of the Berry phase theory~\cite{D-Xiao2010} and Kubo formula~\cite{Kubo1957}, the intrinsic AHC and ANC were expressed as~\cite{WX-Feng2016}
	\begin{eqnarray} 
		\sigma_{ij} &=& -\dfrac{e^{2}}{\hbar}\sum_{n}\int\frac{d^{3}k}{(2\pi)^{3}}\Omega^{n}_{ij}(\bm{k})w_{n}(\bm{k}), \label{eq:IAHC} \\
		\alpha_{ij} &=& -\dfrac{e^{2}}{\hbar}\sum_{n}\int\frac{d^{3}k}{(2\pi)^{3}}\Omega^{n}_{ij}(\bm{k})W_{n}(\bm{k}) , \label{eq:IANC} 
	\end{eqnarray}
	respectively, in which $\Omega^{n}_{ij}(\bm{k})$ is the band-resolved Berry curvature
	\begin{equation}\label{eq:BerryCur}
		\Omega^{n}_{ij}(\bm{k}) = -\sum_{n^{\prime} \neq n}\frac{2\mathrm{Im}[\left\langle \psi_{n\bm{k}}\right|\hat{v}_{i}\left| \psi_{n^{\prime}\bm{k}} \right\rangle \left\langle \psi_{n^{\prime}\bm{k}}\right|\hat{v}_{j}\left|\psi_{n\bm{k}} \right\rangle]}{\left(\omega_{n^{\prime}\bm{k}}-\omega_{n\bm{k}}\right)^{2}}.
	\end{equation}
	
	Here, $\{i,j\}=\{x,y,z\}$ denote the Cartesian coordinates, $\hat{v}_{i, j}$ are velocity operators, and $\psi_{n\bm{k}}$ ($\hbar\omega_{n\bm{k}}=\varepsilon_{n\bm{k}}$) is the eigenvector (eigenvalue) at band index $n$ and momentum $\bm{k}$.  The weighting factors $w_{n}(\bm{k})$ and $W_{n}(\bm{k})$ in Eqs.~\eqref{eq:IAHC} and~\eqref{eq:IANC} were written as
	\begin{eqnarray}
		w_{n}(\bm{k})  &=&  f_{n}(\bm{k}), \label{eq:wnT} \\ 
		W_{n}(\bm{k}) &=& -\dfrac{1}{eT} [(\varepsilon_{n\bm{k}}-\mu) f_{n}(\bm{k}) + \nonumber \\ 
		& & k_{B}T \textrm{ln}(1+e^{-(\varepsilon_{n\bm{k}}-\mu)/k_{B}T})], \label{eq:WnT}
	\end{eqnarray}
	where $f_{n}(\bm{k})=1/[\text{exp}((\varepsilon_{n\bm{k}}-\mu)/k_{B}T)+1]$ is the Fermi-Dirac distribution function, $T$ is temperature, $\mu$ is chemical potential, and $k_{B}$ is the Boltzmann constant. The well converged AHC and ANC were obtained by integrating the Berry curvature and weighting factors over the entire Brillouin zone using a dense $k$-mesh of 200$\times$200$\times$200 points.
	
	To check the validity of our first-principles calculations, we calculated the ANC in several representative magnets, which are listed in Table~\ref{tab2}.  Our results fit well with the previous theoretical and experimental data.


%

\end{document}